\documentclass[prm,twocolumn,showpacs,amsmath,amssymb,superscriptaddress,aps]{revtex4-1}

\usepackage{graphicx}
\usepackage{bm}
\usepackage{physics}
\usepackage{multirow}
\usepackage{hyperref}
\hypersetup{pdfnewwindow=true, colorlinks=true,
 linkcolor=blue, anchorcolor=blue,
 citecolor=blue, filecolor=blue,
 menucolor=blue, urlcolor=blue}

\usepackage{dcolumn}
\newcolumntype{d}[1]{D{.}{.}{#1}}

\begin{document}

\title{What can one learn about material structure given a single first-principles calculation?}

\author{Nicholas Rajen}
\affiliation{Materials Science and Engineering, University of California Riverside, Riverside, CA 92521, USA}
\author{Sinisa Coh}
\affiliation{Materials Science and Engineering, University of California Riverside, Riverside, CA 92521, USA}
\affiliation{Mechanical Engineering, University of California Riverside, Riverside, CA 92521, USA}

\date{\today}

\begin{abstract}
We extract a variable $X$ from electron orbitals $\Psi_{n\bf{k}}$ and energies $E_{n\bf{k}}$ in the parent high-symmetry structure of a wide range of complex oxides: perovskites, rutiles, pyrochlores, and cristobalites.  Even though calculation was done only in the parent structure, with no distortions, we show that $X$ dictates material's true ground state structure.   We propose using Wannier functions to extract concealed variables such as $X$ both for material structure prediction and for high-throughput approaches.
\end{abstract}

\maketitle

\section{Introduction and motivation}
There is a growing need for machines to learn about matter from results of a large set of first-principles computer calculations.  However, a single such calculation produces $\sim 10^6$--$10^9$ bits of information and it is unclear how machine can make use of them.\cite{curtarolo2013,Ghiringhelli2015}  Therefore, so far, machines set aside most of these bits --- such as those describing electron orbitals $\Psi_{n \bf{k}}$ --- and try to learn from the simplest calculated quantities such as the total energy $E^{\rm tot}$ or similar.

One such example where machines are learning from $E^{\rm tot}$ are methods used for predicting positions of atoms in a yet unsynthesized material.  These methods look for a set of atom positions $\xi$ that minimize the total energy $E^{\rm tot} ( \xi )$ of the material.  Total energy of a periodic solid can be computed from first-principles using various approximations to the density functional theory.\cite{Hohenberg1964,Kohn1965}  However,  minimization of $E^{\rm tot} (\xi)$ still remains an unsolved problem as $\xi$ is a vector in a very highly dimensional space (it has $\approx 3N$ dimensions and $N$ is a number of atoms in the material).  Nevertheless, this problem can be addressed heuristically using machine learning techniques such as evolutionary algorithms\cite{Lyakhov2013} or particle swarm optimization.\cite{Wang2012}  Broadly speaking these methods first use $E^{\rm tot}$ calculated for a wide range of different candidate structures $\xi_1, \xi_2,...\xi_m$, but the same chemical composition, to learn about the underlying interactions in that material.  Next, given this knowledge, the machine makes an informed guess for the next structure $\xi_{m+1}$ and the process repeats until an optimal structure is found.  Clearly, if one used in this process information contained in electron orbitals $\Psi_{n\bf{k}}$, and not only $E^{\rm tot}$, one could make a more informed guess of structure $\xi_{m+1}$.

The need for a machine to learn in the context of materials science is also relevant for the so-called high-throughput approaches such as the materials project,\cite{Jain2011} the aflow,\cite{Curtarolo2012} the oqmd,\cite{Saal2013}, aiida,\cite{aiida} or the nomad\cite{nomad} materials databases.  While in the structure prediction problem one considers a single chemical composition at the time, in the high-throughput approach one wishes to learn about materials with a wide range of chemical compositions.  Since these databases contain $\sim10^5$--$10^6$ materials their total information content, if one were to store electron orbitals, is about $10^{14}$--$10^{15}$~bits.

In this paper we do not focus on what or how machines can learn from total energy $E^{\rm tot}$.  Instead, the goal of this paper is to construct a descriptor of electron orbitals $\Psi_{n \bf{k}}$ and eigenenergies that can be used for learning in structure prediction and high-throughput approaches.  As a proof of principle, we construct here a descriptor  --- denoted as $X$ --- that is strongly correlated with the preferred crystal structure of a material, as described later.  Since in our proof of principle work $X$ turns out to be a single number, one can establish a correlation between $X$ and structure just by inspection, without using machine learning.  However, for materials with lower symmetry or in the cases of other properties of interest (i.e., not crystal structure) similarly constructed descriptors of electron orbitals will correspond to more than one number and one would therefore have to use machine learning.

Crystal structures of materials can be divided somewhat loosely into families of structures\cite{Muller2011} based on polyhedral units present in the structure and their connectivity.  Each structure family is derived from a simple high-symmetry structure also called parent, aristotype, or prototype structure.  Remaining structures in the family are then derived from the parent by either displacing or substituting atoms.\cite{Megaw1973,Barnighausen1980}  These derivative structures are also called hettotypes.  When structures are related to each other by distortion that does not preserve polyhedral units or their connectivity, we refer to them only as polymorphs.    

This paper is structured as follows.  In Sec.~\ref{sec:approach} we describe our approach and in Sec.~\ref{sec:results} we present and discuss our results.  A comparison of our results with atomic descriptors  (Pettifor maps) is done in Sec.~\ref{sec:comparison}.
  
\begin{figure*}[!t]
\centering
\includegraphics[width=7in]{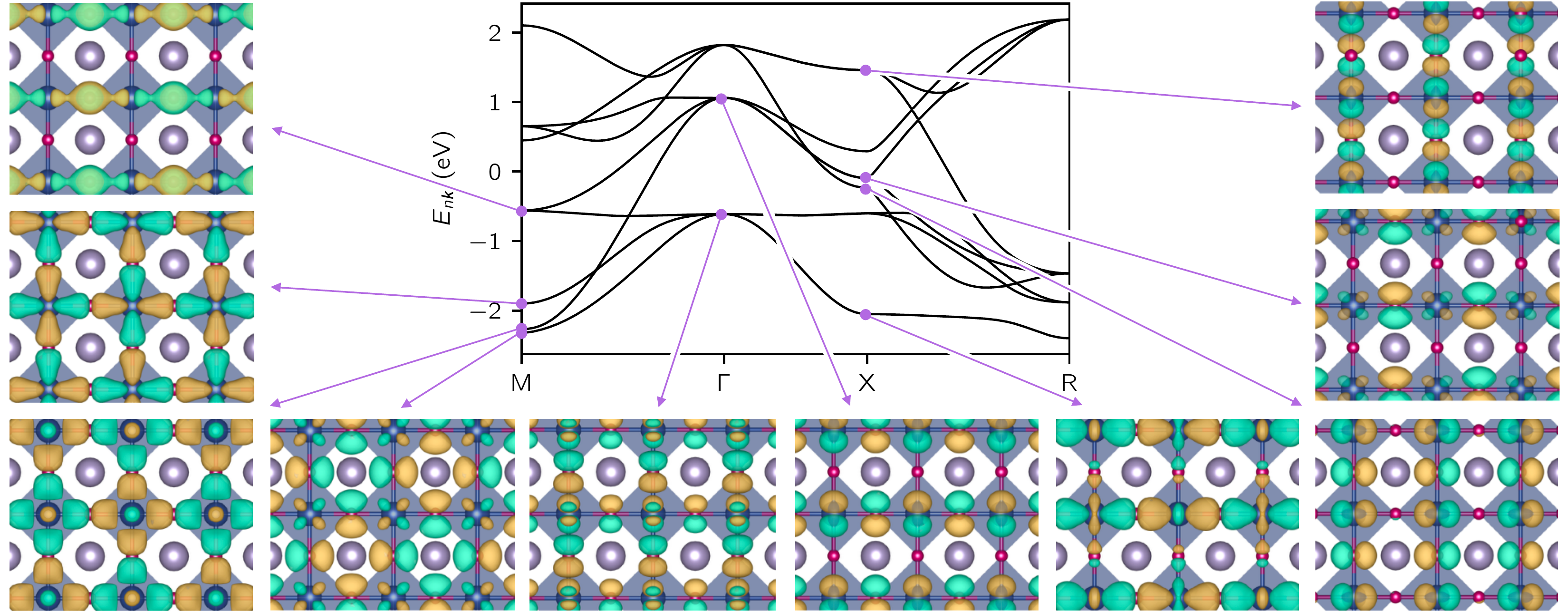}
\caption{\label{fig:wfc} Selection of everal electron orbitals $\Psi_{n \bm{k}}$ in the high-symmetry cubic phase of SrTiO$_3$ perovskite. Yellow/green: positive/negative isosurfaces of electron orbitals. Blue: Ti--O octahedra.  Red: oxygen atoms.}
\end{figure*}

\begin{figure*}[!t]
\centering
\includegraphics[width=7in]{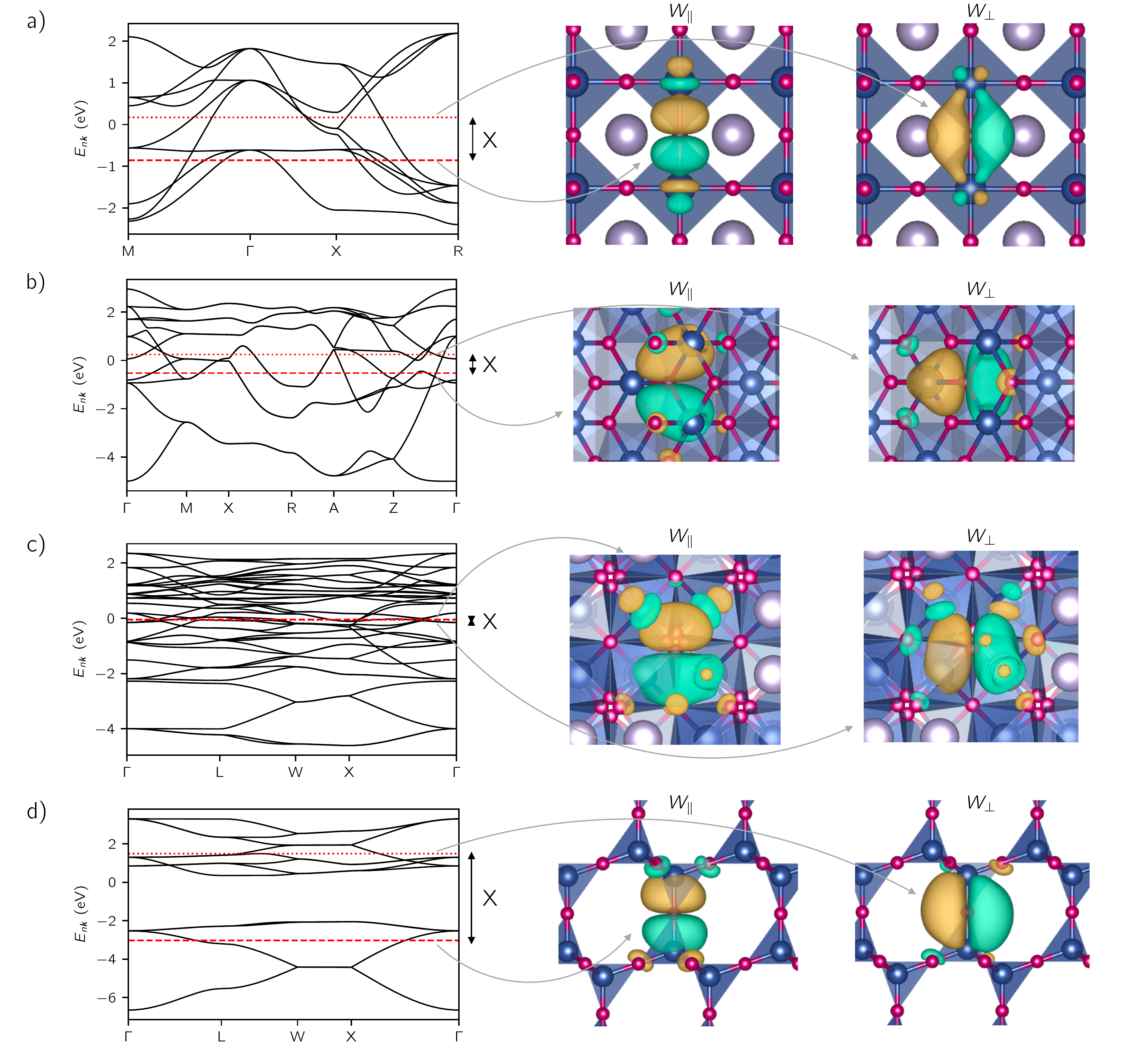}
\caption{\label{fig:allx} Localized Wannier functions (right) and band structure (left) for representatives of all four classes of compounds we studied [SrTiO$_3$ perovskite (a), SnO$_2$ rutile (b), Y$_2$Sn$_2$O$_7$ pyrochlore (c), and SiO$_2$ cristobalite (d)].  Yellow/green: positive/negative isosurfaces of Wannier functions $W_{\perp}$ and $W_{\parallel}$.  Dotted/dashed red line: $\mel{W_{\perp}}{H}{W_{\perp}}$ and $\mel{W_{\parallel}}{H}{W_{\parallel}}$.  The difference is indicated with a black arrow (X). }
\end{figure*}

\begin{figure*}[!t]
\centering
\includegraphics[width=7in]{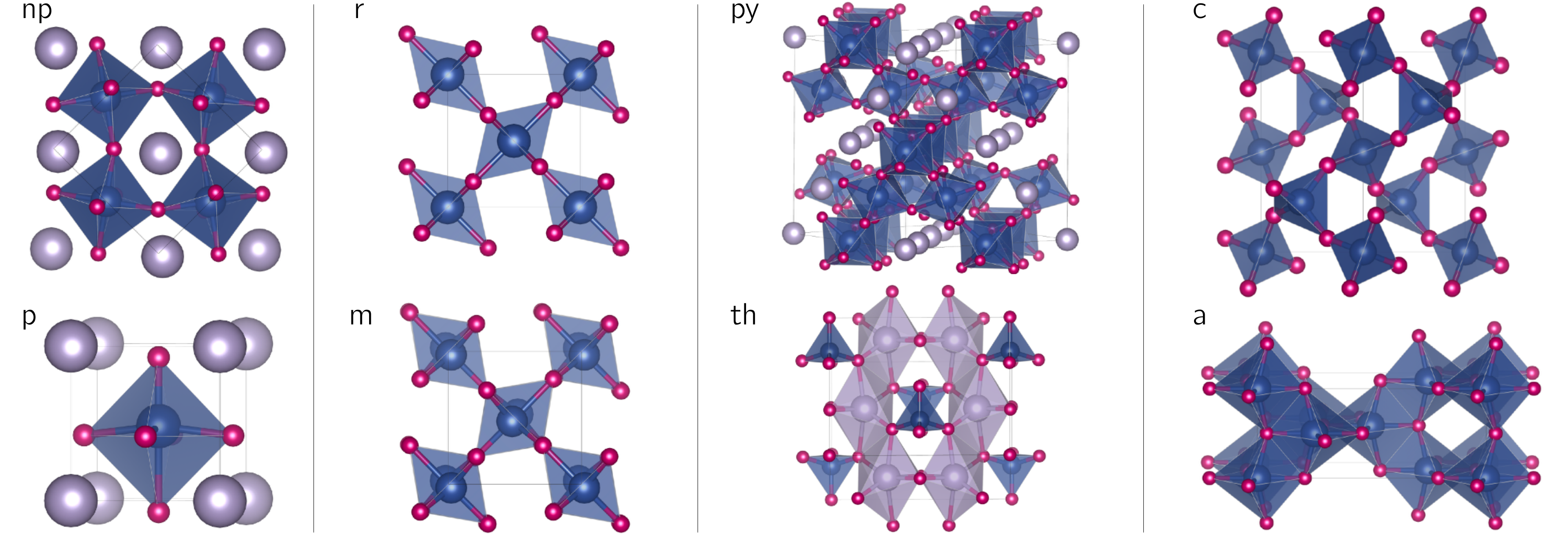}
\caption{\label{fig:str} Crystal structures of the four family classes we considered: perovskites (np and p), rutiles (r and m), pyrochlores (py and th), and cristobalites (c and a).  Meaning of symbols is the same as in Table~\ref{tab:deltaE}.}
\end{figure*}

\section{Approach}
\label{sec:approach}
In what follows we present a general approach to extract a variable $X$ --- given electron orbitals $\Psi_{n \bf{k}}$ and eigenenergies of a parent structure alone --- that dictates which structure is a true ground state structure of that material.  Since such a variable knows about material's structure (i) it has the potential to be used in machine learning approaches described earlier.  To that end, a useful variable should also satisfy two additional criteria: (ii) $X$ should be represented with as few numbers as possible, so that learning is as efficient as possible; and (iii) $X$ should be calculable automatically for a given material without any input from the user, so that it can be used by a machine.

To satisfy the criterion (i) we should ideally extract $X$ in some way from the fundamental variable of the problem.  In the case of quantum mechanical descriptions of matter, the fundamental variable is the electron many-body wave function $\widetilde{\Psi}$.  We note that the many-body wave function $\widetilde{\Psi}$ is formally not accessible in the commonly used Kohn-Sham\cite{Kohn1965} scheme for the density functional theory.  Therefore we instead use here the so-called Kohn-Sham orbitals $\Psi_{n \bm{k}}$ (and corresponding eigenenergies $E_{n \bm{k}}$) as they are typically reasonably good approximations\cite{Hybertsen1986} of the quasiparticle wave function and energy (here $n$ is a band index while $\bm k$ is a crystal momentum).  Figure~\ref{fig:wfc} shows some of the Kohn-Sham orbitals for one of the materials we studied in this work (SrTiO$_3$ cubic perovskite).

Orbitals $\Psi_{n \bm{k}}$ themselves are not a good choice of $X$ as they do not satisfy criterion (ii).  Namely, even in the simplest calculations (say, cubic silicon), a machine needs $\sim 10^6$ bits to describe $\Psi_{n \bm{k}}$ reasonably accurately.  As discussed later, we circumvent this difficulty by defining $X$ in terms of the Wannier functions,\cite{Marzari2012} as they provide a compact and faithful representation of $\Psi_{n \bm{k}}$ (see Fig.~\ref{fig:allx}).  Wannier functions can be computed without user input\cite{Mustafa2015} so our approach satisfies condition (iii) as well.

To demonstrate the generality of our approach we consider here a diversified set of four structure families: {\it AB}O$_3$ perovskites, {\it A}O$_2$ rutiles, {\it A}$_2${\it B}$_2$O$_7$ pyrochlores, and {\it A}O$_2$ cristobalites for a range of A and B anions.  We chose these four families because they display different bonding environments containing both octahedral and tetrahedral oxygen units, as well as different connectivity of these units (see Fig.~\ref{fig:str}).  The main result of this paper is that we find that our general approach yields an energy scale $X$ that is very well correlated with preferred ground state structure.  We studied 64 representative compounds: 17 perovskites, 16 rutiles, 18 pyrochlores, and 13 cristobalites.

The variable $X$ is constructed in five steps.  {\it First}, we relax the structure of each compound while preserving the symmetry of the parent structure (aristotype).\footnote{We use generalized gradient approximation\cite{Perdew1996} to the density functional theory within the quantum-espresso package\cite{Giannozzi2009,PhysRevB.88.085117,Garrity2014} Wannier functions were constructed using Wannier90 package.\cite{Mostofi2008685}}

{\it Second}, we identify a set of electron bands of interest given the band structure from the first step.  In our case we simply take as bands of interest a complex of bands with a dominant oxygen $p$-like character, as they are the most dispersive and thus contain the most information about the inter-atomic interactions in the material.  Oxygen bands are also fully occupied and isolated in these materials, which simplifies our Wannier function based analysis.  Figure~\ref{fig:allx} shows bands of interest in the representative case of each of the four families we studied. 

{\it Third}, we convert electron orbitals $\Psi_{n \bm{k}}$ of bands selected in step two into a basis of maximally localized Wannier functions $W_{n}$.\cite{Wannier1937,Marzari1997,Marzari2012}  This procedure significantly simplifies our analysis as Wannier functions are typically strongly localized on a single atom\cite{Marzari1997,Brouder2007} and are therefore more convenient descriptors of the chemical environment of a material. Given a set of extended periodic orbitals $\Psi_{n \bm{k}}$ Wannier functions $W_{n}$ are defined as,
\begin{align}
W_n (\bm{r}) = \frac{1}{N_{\bm{k}}} \sum_k U_{nm\bm{k}} \Psi_{m \bm{k}} (\bm{r})
\label{eq:wan}
\end{align}
where $N_{\bm{k}}$ is a total number of  $k$-points.  While unitary matrices $U_{nm\bm{k}}$ are in principle arbitrary, in the case of maximally localized Wannier functions they are chosen so that the resulting function $W_n(\bm{r})$ is as localized in real space as possible.\cite{Marzari1997}
We stress that basis change in Eq.~\eqref{eq:wan} is exact in the sense that the vector space spanned by $W_n$ (and its periodic images) is the same as the space spanned by $\Psi_{m \bm{k}}$.  However, $W_n$ is a more convenient object to study than $\Psi_{n \bm{k}}$ since a single function $W_n$ contains the same information as an entire band of Bloch functions $\Psi_{n \bm{k}}$ (there is one function for each $\bm k$-point).  Therefore any information contained in bands is also contained in the corresponding Wannier functions.  One of the Wannier functions for one of the parent structures (perovskites) is denoted in Fig.~\ref{fig:allx}~(a) as $W_{\parallel}$.  This Wannier function corresponds to the oxygen $p$-like function oriented along the B--O--B line (here B is the anion in the ABO$_3$ perovskite that is in the center of the oxygen octahedron).  The remaining two functions ($W_{\perp}$) centered on the same oxygen atom are perpendicular to the B--O--B line [see also Fig.~\ref{fig:allx}~(a)].  In the remaining three families (rutile, pyrochlore, cristobalite) we also find that Wannier function on each oxygen atom follow the same logic: there is a singlet orbital (denoted again as $W_{\parallel}$) and a degenerate pair of orbitals ($W_{\perp}$).  These Wannier functions are shown in  Figs.~\ref{fig:allx}~(b), (c), and (d).

{\it Fourth}, we construct the representation of the Hamiltonian operator $H$ in the Wannier basis from the previous step,
\begin{equation}
H_{nm \bm{R}} = \mel{W_{n}}{H}{W_{m \bm{R}}}.
 \label{eq:WFHam}
\end{equation}
Here  $W_{m \bm{R}}$ is defined as $W_{m}$ translated by a lattice vector $\bm R$,
$
W_{m \bm{R}} (\bm{r}) = W_{m} (\bm{r} - \bm{R}).
$
It is straightforward to show that for structural properties we need not consider $H_{nm \bm{R}}$ when either $n \neq m$ or $\bm{R} \neq \bm{0}$.  These terms are often referred to as hopping integrals.  Hopping integrals do not contribute to the integrated band energy $\int_{\rm BZ} E_{n \bm{k}} d\bm{k}$ of a fully occupied band and therefore they don't have the learning potential for determining ground state structure.\footnote{Total energy $E^{\rm tot}$ in the Kohn-Sham framework does not depend only on $\int_{\rm BZ} E_{n \bm{k}} d\bm{k}$ but also on additional terms that are explicit functions of electron density $n(\bm{r})$.  However, density can be computed from the Wannier function as $n(\bm{r}) = \sum_{n \bm{R}} | W_n (\bm{r})|^2$. Therefore, in some cases it is possible that one would have to include a descriptor for the shape of the Wannier function in order to learn about the material structure.  That is not the case here as we consider materials in their parent structure, so oxygen Wannier function have nearly constant shape regardless of anion {\it A} or {\it B}.} Therefore, we are left with $H_{nm \bm{R}}$ when both $n=m$ and $\bm{R}=\bm{0}$.  The matrix element $H_{nn \bm{0}}$ is usually referred to as the onsite energy of the $n$-th Wannier function. The absolute value of onsite energy is ill-defined for a periodic solid as one can change its value by adding an arbitrary constant $C$ to the Hamiltonian of the periodic solid: $H \rightarrow H+C$.  However, changing $H$ to $H+C$ changes onsite energies of all Wannier functions by the same amount.  Therefore, even though absolute values are ill-defined, the differences of onsite energies,
\begin{align}
H_{nn \bm{0}} - H_{mm \bm{0}}
 \label{eq:diffh}
\end{align}
are well defined.

{\it Fifth}, we use symmetry to find all distinct values of $H_{nn \bm{0}} - H_{mm \bm{0}}$ from the previous step.  Since the symmetry of all four parent structures is high, we find that in all four families there remains only one distinct numerical value in which $n$ corresponds to $W_{\perp}$ and $m$ to $W_{\parallel}$.  This difference we denote as $X$,
\begin{align}
X = \mel{W_{\perp}}{H}{W_{\perp}} - \mel{W_{\parallel}}{H}{W_{\parallel}}.
 \label{eq:ediff}
\end{align}
As an example, Fig.~\ref{fig:allx}~(a) shows a complex of oxygen $p$-like electron bands (black) in SrTiO$_3$ along with calculated $\mel{W_{\perp}}{H}{W_{\perp}}$ and $\mel{W_{\parallel}}{H}{W_{\parallel}}$ (dotted/dashed red line).  The black arrow indicates their difference ($X$).  The numerical value of $X$ for all materials we studied is provided in Tab.~\ref{tab:deltaE}. 

We stress here that the energy scale $X$ can't be inferred from the electron band-energies  $E_{n \bm{k}}$ alone. Instead, one also needs to use wave functions $\Psi_{n\bm{k}}$ in the construction.  In addition, $X$ is gauge-dependent as a different choice of relative phases of wave functions will lead to a different Wannier function and thus different $X$.  For example, one can show that a simple unitary rotation that rotates $W_{\parallel}$ and $W_{\perp}$ will produce two Wannier function with the difference in onsite energy having any value between $+X$ and $-X$, including zero.  However, this procedure will increase the spread of the Wannier functions and will reduce their symmetry.  Therefore, in this work we constructed Wannier functions for each family structure in a consistent way that respects the symmetry of the parent phase.

\begin{table*}
\caption{\label{tab:deltaE}
Variable $X$ calculated in a high-symmetry parent state correlates well with the preferred ground state structure (third subcolumn).  The meanings of the crystal structure abbreviations are given at the bottom of the table.}
\begin{ruledtabular}
\begin{tabular}{ld{2.2}c ld{2.2}c ld{2.2}c ld{2.2}c}
\multicolumn{3}{l}{Perovskite} &
\multicolumn{3}{l}{Rutile} &
\multicolumn{3}{l}{Pyrochlore} &
\multicolumn{3}{l}{Cristobalite} \\
\cline{1-3} \cline{4-6} \cline{7-9} \cline{10-12}
\multirow{2}{*}{{\it AB}O$_3$}              &  \multicolumn{1}{c}{X} &  \multirow{2}{*}{Struc.} &        
\multirow{2}{*}{{\it A}O$_2$}               &  \multicolumn{1}{c}{X} &  \multirow{2}{*}{Struc.} &        
\multirow{2}{*}{{\it A}$_2${\it B}$_2$O$_7$}&  \multicolumn{1}{c}{X} &  \multirow{2}{*}{Struc.} &
{\it A}O$_2$                                &  \multicolumn{1}{c}{X} &  \multirow{2}{*}{Struc.} \\
  &  \multicolumn{1}{c}{(eV)} &  &             
  &  \multicolumn{1}{c}{(eV)} &  &        
  &  \multicolumn{1}{c}{(eV)} &  &
  {\it AB}O$_4$   &  \multicolumn{1}{c}{(eV)} &  \\
\cline{1-3} \cline{4-6} \cline{7-9} \cline{10-12}
CaZrO$_3$ & 0.47 &  np    &	ReO$_2$ &  -0.27   & m      & 	Mg$_2$P$_2$O$_7$  & -1.47 &  th	&	 InSbO$_4$ & 3.62  & a  	\\
SrZrO$_3$ & 0.57 &  np    &	WO$_2$  &  -0.26   & m      & 	Mg$_2$As$_2$O$_7$ & -0.96 &  th	&	 SnO$_2$   & 3.68  & a  	\\
CaHfO$_3$ & 0.66 &  np    & MoO$_2$ &  -0.18   & m      & 	Y$_2$Si$_2$O$_7$  & -0.73 &  th	&	 PbO$_2$   & 3.80  & a  	\\
PbZrO$_3$ & 0.70 &  np    &	TcO$_2$ &  -0.09   & m      & 	In$_2$Si$_2$O$_7$ & -0.60 &  th	&	 AlSbO$_4$ & 3.95  & a  	\\
BaZrO$_3$ & 0.76 &  np    &	NbO$_2$ &  -0.06   & n      & 	Cd$_2$V$_2$O$_7$  & -0.59 &  th	&	 GaSbO$_4$ & 3.98  & a  	\\
SrHfO$_3$ & 0.76 &  np    &	VO$_2$  &  0.11    & m      & 	Sc$_2$Si$_2$O$_7$ & -0.54 &  th	&	 InPO$_4$  & 4.00  & c  	\\
PbHfO$_3$ & 0.90 &  np    & TiO$_2$ &  0.17    & r      & 	In$_2$Ge$_2$O$_7$ & -0.31 &  th	&	 InAsO$_4$ & 4.15  & c  	\\
CdTiO$_3$ & 0.90 &  p     & CrO$_2$ &  0.22    & r      & 	La$_2$Sn$_2$O$_7$ & -0.27 &  py	&	 GaPO$_4$  & 4.43  & c  	\\
CaTiO$_3$ & 0.91 &  np    &	MnO$_2$ &  0.27    & r      & 	Sc$_2$Ge$_2$O$_7$ & -0.22 &  th	&	 AlPO$_4$  & 4.47  & c  	\\
BaHfO$_3$ & 0.96 &  np    &	RuO$_2$ &  0.27    & r      & 	Y$_2$V$_2$O$_7$   & -0.18 &  py	&	 SiO$_2$   & 4.52  & c  	\\
SrTiO$_3$ & 1.03 &  p     &	OsO$_2$ &  0.39    & r      & 	Bi$_2$Ti$_2$O$_7$ & -0.14 &  py	&	 GaAsO$_4$ & 4.54  & c  	\\
PbTiO$_3$ & 1.20 &  p     &	IrO$_2$ &  0.45    & r      &	Y$_2$Mo$_2$O$_7$  & -0.10 &  py	&	 AlAsO$_4$ & 4.56  & c  	\\
BaTiO$_3$ & 1.25 &  p     &	SnO$_2$ &  0.77    & r      & 	Y$_2$Ti$_2$O$_7$  & -0.04 &  py	&	 GeO$_2$   & 4.67  & c  	\\
NaNbO$_3$ & 1.31 &  p     &	PbO$_2$ &  0.82    & r      & 	La$_2$Hf$_2$O$_7$ & -0.04 &  py	&	           &       &    	\\
KNbO$_3$  & 1.45 &  p     &	SiO$_2$ &  1.00    & r      & 	La$_2$Pb$_2$O$_7$ & -0.03 &  py	&	           &       &    	\\
LaAlO$_3$ & 1.54 &  np    & GeO$_2$ &  1.06    & r      &	Y$_2$Sn$_2$O$_7$  & -0.02 &  py	&	           &       &    	\\
KTaO$_3$  & 1.66 &  p     &	        &          &        & 	La$_2$Zr$_2$O$_7$ &  0.01 &  py	&	           &       &    	\\
          &      &        &	        &          &        &	Bi$_2$Hf$_2$O$_7$ &  0.02 &  py	&	           &       &    	\\
\cline{1-3} \cline{4-6} \cline{7-9} \cline{10-12}
\multicolumn{3}{l}{np = non-polar hettotype} & 
\multicolumn{3}{l}{r = rutile hettotype} & 
\multicolumn{3}{l}{py = pyrochlore} &
\multicolumn{3}{l}{c = cristobalite} \\
\multicolumn{3}{l}{p = polar hettotype} & 
\multicolumn{3}{l}{m = manganite hettotype} & 
\multicolumn{3}{l}{th = thortveitite} &
\multicolumn{3}{l}{a = anatase } \\
\multicolumn{3}{l}{} & 
\multicolumn{3}{l}{n = NbO$_2$ hettotype} & 
\multicolumn{3}{l}{} &
\multicolumn{3}{l}{} \\
\end{tabular}
\end{ruledtabular}
\end{table*}

\section{Results}
\label{sec:results}
Table~\ref{tab:deltaE} contains calculated values of $X$ for the 64 compounds we studied.  As can be seen from the table, $X$ is well correlated with the preferred ground state structure.  We discuss now  all four structure families in more detail.

{\it Perovskites} are one of the most studied complex oxides.  The numerical values of $X$ reported in Tab.~\ref{tab:deltaE} for perovskites was calculated in the parent cubic phase with space group Pm$\bar{3}$m.   As can be seen from the table, $X$ is positive for all perovskites and it ranges from $0.47$~eV to $1.66$~eV. There is a large number of structures (hettotypes) that derive from the cubic perovskite parent structure.  Some of these structures are polar (denoted as p in Tab.\ref{tab:deltaE} and Fig.~\ref{fig:str})  while others are non-polar and have rotated oxygen octahedra (np).\cite{Glazer:a09401, King-Smith1994, Woodward1997, Benedek2013} Our analysis shows that perovskites with $X$ less than $\sim 1$~eV tend to condense into a non-polar state (np) while those with $X$ larger than $\sim 1$~eV condense into a polar state (p).  The exception is CdTiO$_3$ which is polar but has $X=0.90$~eV and LaAlO$_3$ which is non-polar but has $X=1.54$~eV.  While the value of $X$ for CdTiO$_3$ is near the polar--non-polar boundary, LaAlO$_3$ is not.  The anomalous value of $X$ for LaAlO$_3$ likely occurs because its perovskite phase is degenerate with another structure.\cite{osti_1203574}

We note here that the importance of local interactions in perovskites has been discussed earlier in Refs.~\cite{Cohen1992, Bersuker1995, Woodward1997, Palmer1997, Kolezynski2005}. However, we are unaware of any other work in which ground state density functional calculation of a high-symmetry perovskite alone can be used to infer its low-symmetry ground state structure.

{\it Rutiles} are another common structure of complex oxides.\cite{Baur2007}  Unlike perovskites, rutiles have a somewhat more complicated structure as their octahedral units are both corner and edge shared. The numerical values of $X$ reported in Tab.~\ref{tab:deltaE} for rutiles was calculated in the parent rutile structure (r) with tetragonal space group P4$_2$/mnm.  As can be seen from the table compounds with $X$ larger than $0.17$~eV remain in the rutile phase, while those with smaller $X$ distort into a manganite (m) or NbO$_2$-type (n) structure.\cite{Baur2007} Both of these structures contain off-centered {\it A} anions and deformed oxygen octahedra. The more common, manganite structure, is monoclinic with space group P2$_1$/c while the structure unique to NbO$_2$ is tetragonal I4$_1$/a.  In rutiles calculated value of $X$ ranges between $-0.27$~eV and $1.06$~eV.

{\it Pyrochlores} have the most complex structure among the materials we studied. Variable $X$ was computed in the cubic pyrochlore (py) state with space group Fd$\bar{3}$m.  We find that compounds with $X$ above $-0.2$~eV remain in the pyrochlore state at their ground state while others form into thortveitite (th)\cite{Brisse1972} polymorph\cite{family} with monoclinic space group C2/m.  The only outlier we found is La$_2$Sn$_2$O$_7$ as it has $X=-0.27$~eV but, as far as we are aware, it remains in a pyrochlore structure at a ground state.  

{\it Cristobalite} is a well known structure in which anions are surrounded with oxygen tetrahedra, unlike the other three families in which oxygen atoms form octahedral units.   The numerical values of $X$ reported in Tab.~\ref{tab:deltaE} for cristobalites were calculated in the parent cubic idealized cristobalite structure (space group F$\bar{4}$3m).  We again find a strong correlation between $X$ and the preferred ground state structure.  Compounds with $X$ larger than $\sim 4$~eV remain cristobalite (with small deviations known as $\alpha$ or $\beta$-cristobalite, space groups P4$_1$2$_1$2 and I$\bar{4}$2d).  Those with $X$ less than $\sim 4$~eV are unstable in the cristobalite structure and distort instead into anatase phase (a) with tetragonal space group I4$_1$/amd.\cite{family}  In the cristobalites we studied the calculated values of $X$ range from $3.62$ to $4.67$~eV.

\subsection{Comparison with Pettifor maps}
\label{sec:comparison}
\begin{figure*}[!h]
\centering
\includegraphics[width=5in]{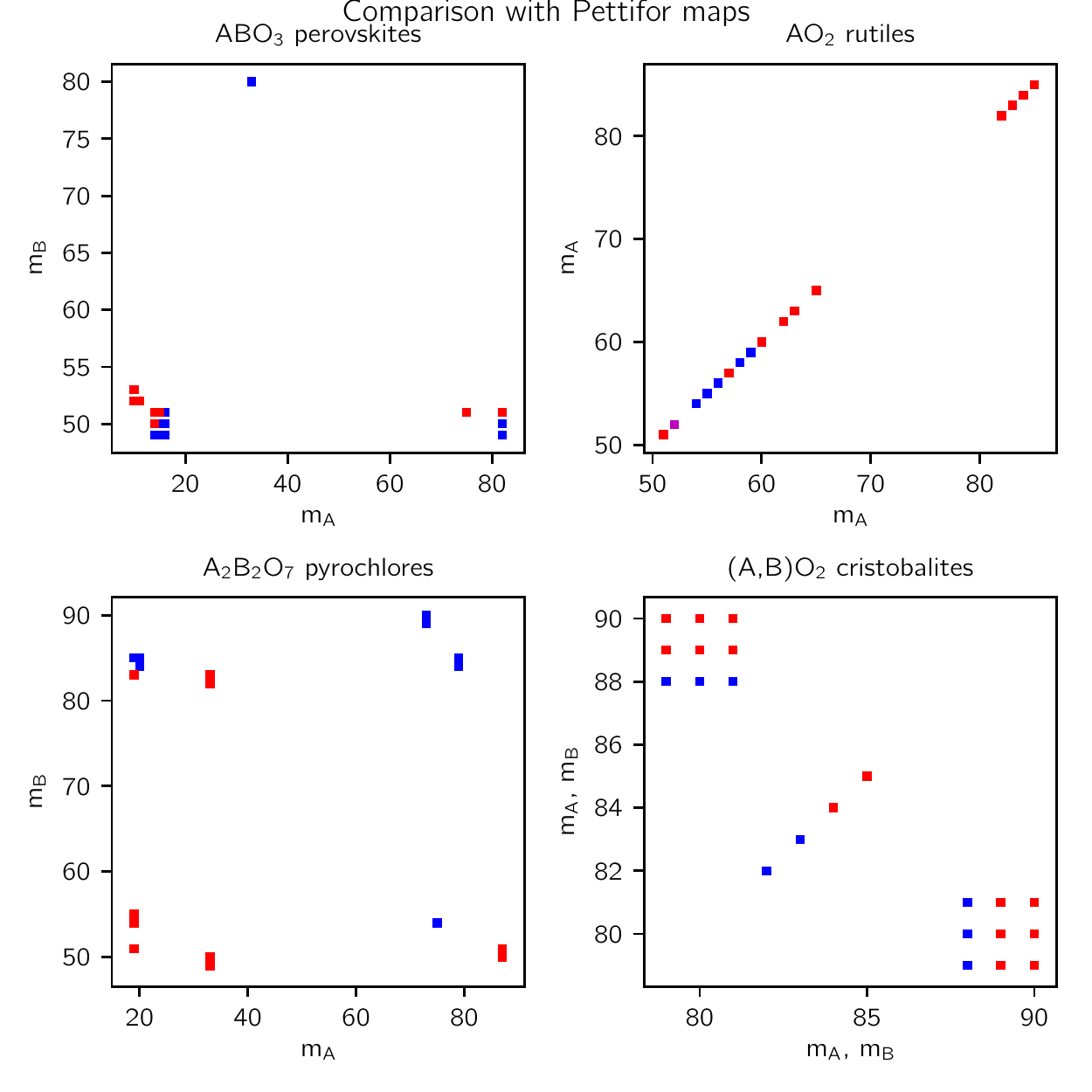}
\caption{\label{fig:pett} Comparison with Pettifor maps.  Red squares correspond to structures p, r, c, py while blue crosses correspond to np, m, a, th.  Conventions for structures are the same as in Table~\ref{tab:deltaE}.}
\end{figure*}

While descriptor $X$ in this work was extracted from a first-principles calculation of a periodic solid, there are several descriptors in use that can be extracted just based on the chemical composition of the solid.  One such example is Pettifor maps\cite{pettifor} which assign a phenomenological chemical scale $\chi$ to each chemical element\cite{pettifor2} in the periodic table.  Figure~\ref{fig:pett} shows such maps for all of the materials studied in this work.  Horizontal and vertical axes in these plots correspond to the ordering of $\chi$ for metal atoms appearing in each compound (this ordering is also called Mendeleev number $m$).

As can be seen from the Fig.~\ref{fig:pett}, Pettifor maps for these materials correlate well with the ground state structure only in the case of cristobalites.  Namely, compounds with smaller $m$ ($\chi$) tend to deform into anatase phase (blue cross symbols) while those with larger $m$ tend to remain in the cristobalite phase (red square symbols).  In the other three classes of materials correlations with Pettifor maps are not as good.  For example, for rutiles we find that those with very large $m$ tend to remain in a rutile phase, but those with lower $m$ can be either rutile or manganite.  In fact, the one with the lowest $m$ (TiO$_2$) is in the rutile phase.  Similarly, no obvious correlation is found for perovskites and pyrochlores.

\section{Conclusion}
\label{sec:conclusion}
Our work shows that it is possible to extract information from a first-principles calculation on a parent phase and correlate it with the preference for a ground state structure.  While few of these correlations can be found using empirical parameters such as Pettifor maps, or ionic radii, our work shows that one can also extract relevant parameters from the fundamental variable in the problem ($\Psi_{n {\bm k}}$).

We expect that one could use this approach not only to characterize Hamiltonian operator $H$ but also any other quantum-mechanical operator, such as electron-light interaction,\cite{ChinShen2017} electron-phonon interaction,\cite{Giustino2007} or other interactions.

\bibliography{pap}

\end{document}